\documentclass[a4paper,3p,11pt,preprint]{elsarticle}

\journal{International Journal of Radiation Oncology, Biology, Physics}

\begin{document}

\begin{frontmatter}

\title{Clinically applicable Monte Carlo-based biological dose optimization for the treatment of head and neck cancers with spot-scanning proton therapy}

\author{H. Wan Chan Tseung\fnref{ref1}}
\ead{wanchantseung.hok@mayo.edu}
\author{J. Ma\fnref{ref1}}
\author{C. R. Kreofsky}
\author{D. Ma}
\author{C. Beltran}

\fntext[ref1]{These authors contributed equally to this work.}
\address{Department of Radiation Oncology, Mayo Clinic, Rochester MN}

\begin{abstract}

{\bf Purpose:} To demonstrate the feasibility of fast Monte Carlo (MC) based inverse biological planning for the treatment of head and neck tumors in spot-scanning proton therapy. 
{\bf Methods:} Recently, a fast and accurate Graphics Processor Unit (GPU)-based MC simulation of proton transport was developed and used as the dose calculation engine in a GPU-accelerated IMPT optimizer. Besides dose, the MC can simultaneously score the dose-averaged linear energy transfer (LET$_d$), which makes biological dose (BD) optimization possible. To convert from LET$_d$ to BD, a simple linear relation was assumed. Using this novel optimizer, inverse biological planning was applied to four patients, including two small and one large thyroid tumor targets, and one glioma case. To create these plans, constraints were placed to maintain the physical dose (PD) within 1.25 times the prescription while maximizing target BD. For comparison, conventional IMRT and IMPT plans were also created using Eclipse (Varian Medical Systems) in each case. The same critical structure PD constraints were used for the IMRT, IMPT and biologically optimized plans. The BD distributions for the IMPT plans were obtained through MC re-calculations. 
{\bf Results:} Compared to standard IMPT, the biologically optimal plans for patients with small tumor targets displayed a BD escalation that was around twice the PD increase. Dose sparing to critical structures was improved compared to both IMRT and IMPT. No significant BD increase could be achieved for the large thyroid case, and when the presence of critical structures mitigated the contribution of additional fields. The calculation of the biological optimized plans can be completed in a clinically viable time (below 30 minutes) on a small 24-GPU system.
{\bf Conclusion:} By exploiting GPU acceleration, MC-based, biologically optimized plans were created for small-target tumor patients. This optimizer will be used in an upcoming feasibility trial on LET$_d$ painting for radio-resistant tumors.

\end{abstract}

\end{frontmatter}


\section{Introduction}

The in-vivo radio-resistance of human tumor cells varies greatly.  Some tumor types have high degrees of radio-resistance both in-vitro and in clinical experience.  These tumors include glioblastoma multiforme \cite{Shu}, poorly differentiated/anaplastic thyroid cancer \cite{Ke}, unresectable salivary tumors \cite{Jakobsson}, certain head and neck sarcomas \cite{Weich1990}, and head and neck squamous cell carcinomas that recur after prior radiotherapy \cite{Weich1988}. The lack of durable radiation response within these tumor types poses a significant clinical challenge. Patients with these tumors have some of the worst disease-free and overall survival statistics among all cancer patients.

Double-strand breaks generated by particle tracks with high Linear Energy Transfer (LET) are thought to be more complex and more difficult to repair \cite{Paganetti1}. The treatment of salivary tumors, thyroid malignancies and glioblastoma with neutrons and carbon ions has demonstrated increased local control when compared with historical controls. For protons, the LET and Relative Biological Effectiveness (RBE) vary along particle trajectories, increasing towards track ends. Although an absolute prediction of RBE values is very difficult, the relationship between RBE and LET can be approximated as linear. Proton treatment plans that exploit this rise in LET to deliver increased biological dose (BD) to the target can potentially result in improved outcomes for radio-resistant tumors.

To estimate BD distributions, one requires the dose-averaged LET (LET$_d$). The most reliable way to calculate LET$_d$ is through Monte Carlo (MC) simulations, since contributions from secondary particles produced in non-elastic proton-nucleus interactions can be significant. Previous authors have employed a variety of methods to elevate the target BD \cite{Fager, Giantsoudi}. BD escalation methods that involve the use of MC techniques are usually of the forward-planning type. Furthermore, they are not clinically viable because CPU-based MC calculations are overly time-consuming. To the best of our knowledge, MC-calculated LET$_d$ distributions have not yet been directly incorporated in inverse planning for proton therapy. Compared to other BD escalation methods, MC-based inverse biological planning is not only expected to be more accurate, but also to produce plans with more conformal and homogeneous BD, and with better normal tissue sparing.

The goal of this work is to demonstrate clinically-applicable, MC-based inverse biological planning in pencil-beam scanning proton therapy. The target BD is escalated in comparison with conventional Intensity Modulated Proton Therapy (IMPT) planning, while keeping the same Organ-At-Risk (OAR) constraints and restricting physical dose (PD) hotspots to within 25\% or so of the prescription dose. Two pieces of software that made such plans calculable in less than 30 minutes are: (a) a GPU-based proton transport MC \cite{Wan}, and (b) a GPU-based IMPT optimizer that uses pre-calculated, spot-wise LET$_d$ and PD maps from the MC \cite{Ma}. The applicability and limitations of the method is illustrated with four cases. The focus is on head-and-neck tumors, although other sites can benefit as well.

\section{Methods and materials}

\subsection{GPU-accelerated MC}

Recently, our group has developed a very fast and accurate proton transport MC for the GPU, with net computational times of $\sim$20s for $1\times 10^7$ proton histories \cite{Wan}. This GPU-based MC includes a Bertini cascade simulation \cite{Wan2} for handling non-elastic interactions on an event-by-event basis, and is thus capable of performing accurate LET$_d$ computations. The MC model for computing the PD was extensively verified with TOPAS version beta-6 \cite{Perl} and Geant4.9.6. The LET$_d$ computations were also compared with TOPAS predictions. As an example, Figure 1 shows the PD and LET$_d$ calculations from TOPAS and our in-house MC for a head-and-neck patient. The 3D-gamma \cite{Low} pass rate at 2\%/2mm was above 98\% for the dose calculation. 

\begin{figure}
\includegraphics[width=\linewidth]{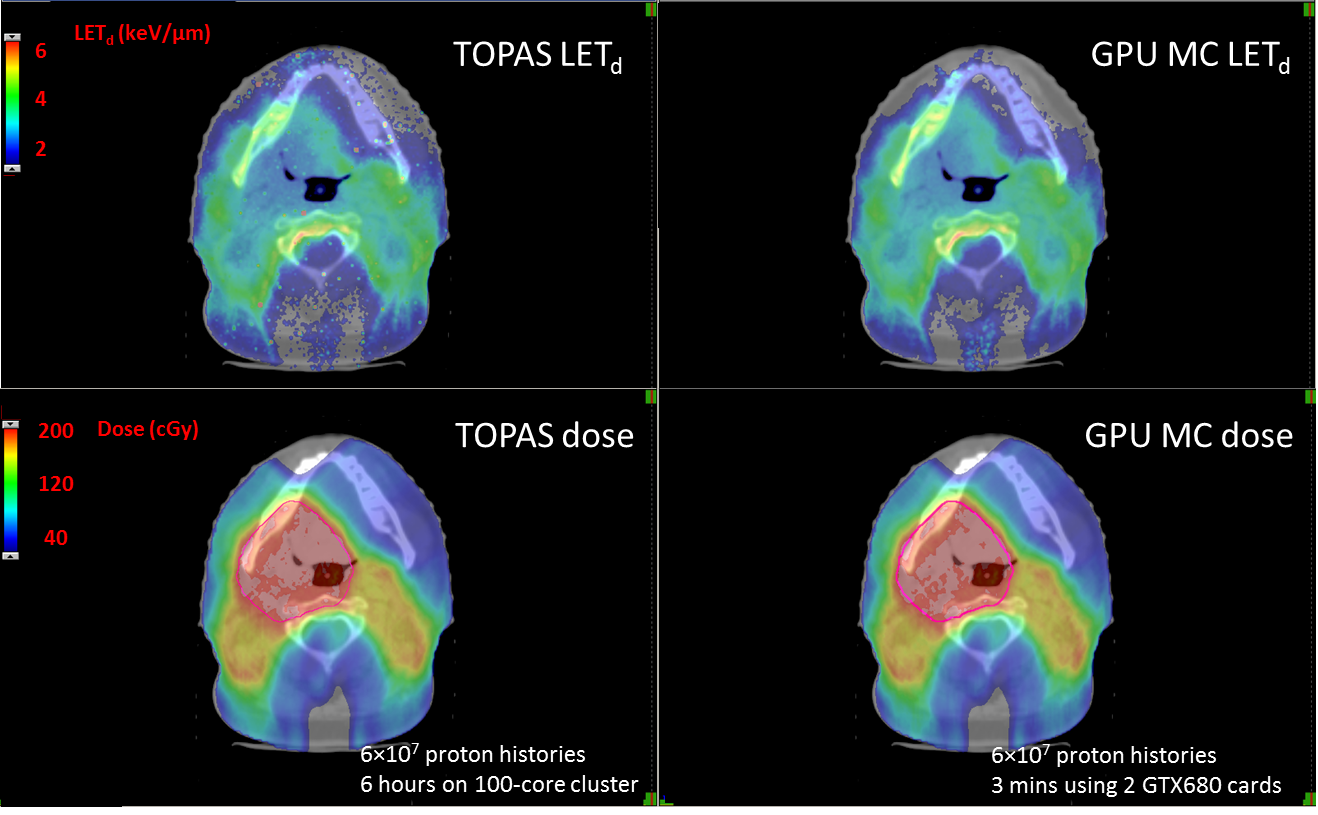}
\caption{TOPAS LET$_d$ and physical dose calculations are shown in the top and bottom left panels, respectively. The corresponding computations from the in-house GPU-based MC are shown on the right.}
\end{figure}

\subsection{GPU-based optimizer}
In addition, our group has built a fast GPU-accelerated IMPT optimizer \cite{Ma}, which uses the previously-described MC as the dose calculation engine. On a 24-GPU system, this optimizer can generate IMPT plans within a clinically viable time frame (i.e. within 30 minutes). Because they are MC-based, these plans are not adversely affected by inaccuracies due to tissue heterogeneities. The IMPT optimizer was slightly modified for the purposes of this work, by allowing the input of LET$_d$ maps, implementing voxel-wise BD estimates (see Equation 1 below), and modifying the cost function. 

\subsection{LET$_d$ to biological dose conversion}
To simplify the BD calculation during the optimization process, a simple linear relationship between LET$_d$ and BD was assumed:
\begin{equation} BD = 1.1\times (a\cdot LET_{d} + b) \end{equation}
where $a$=0.05 $\mu$m/keV and $b$=0.95. PD and LET$_d$ are in units of Gy and keV/$\mu$m, respectively. These coefficients were chosen so that Equation 1 roughly corresponds to a linear-quadratic model with $\alpha/\beta=3$ Gy. With these values of a and b, Equation 1 yields a dependence of RBE on LET$_d$ that is close to previously published models from Carabe et al \cite{Carabe}, Wedenberg et al \cite{Wedenberg} and McNamara et al \cite{McNamara}, which were derived from fits to in-vitro cell survival data. A dose per fraction of 2 Gy was assumed when comparing Equation 1 to these models. The post-optimization BD was evaluated by using the published models instead of Equation 1.  

The term in brackets is the deviation from the conventional $1.1\times PD$. Suggestive evidence for a linear relation between RBE and LET can be found in references \cite{Carabe, Paganetti2}. The same linear relation was naively assumed to be valid for all cell kinds. As discussed below, the choice of $a$ and $b$ values (and hence $\alpha/\beta$) does not affect the ability of the optimizer to converge to a plan with increase target LET$_d$. The assumed value of $\alpha/\beta=3$ Gy is our rough guess for highly radio-resistant tumors.

\subsection{Creation and assessment of biologically-optimized plans}

For each patient in this study, the following three plans were created:
\begin{enumerate}[1.]
\item a photon IMRT plan
\item a conventional two or three-field IMPT plan
\item a five-field biologically optimized plan, using the in-house treatment planning system (TPS) described above. 
\end{enumerate}

The IMRT and IMPT planning were done using a commercially-available TPS (Eclipse, Varian Medical Systems). The number of fields used in the conventional IMPT plans is reflective of our current clinical practice.

To create the biologically optimized plan, an Optimization Target Volume (OTV) \cite{otv}, which is akin to a planning target volume (PTV), was derived from the physician-delineated Clinical Target Volume (CTV), and expanded by 0.5 cm to create a scanning target volume (STV). The MC was then used to place spots inside the STV, with a hexagonal spot spacing of 0.5 cm, and the PD and LET$_d$ maps for each spot was calculated using $10^{5}$ proton histories. To minimize GPU memory use, voxels receiving less than 0.5\% of the maximum dose were ignored \cite{Ma}. Constraints were placed so as to maintain the OTV PD to within $\sim$25\% of the prescription dose, while maximizing the OTV BD. To achieve this, a lower limit was placed on the OTV BD (115\% of prescription dose) and the relative weights of the PD constraint and BD objective were adaptively changed during each optimization step. The biologically-optimized, IMPT and IMRT plans had the same OAR PD constraints. The plan that best satisfies the above requirements is the `biologically optimal' plan.

It should be stressed that constraints were not set on the BD for OARs, since absolute BD values are currently not reliably calculable. Even if they were, relating the high-LET proton BD to past photon-based experience is not trivial. In this work, the proton PD constraints for OARs in plans 2 and 3 were derived from photon dose constraints, assuming an RBE of 1.1. 

The IMPT plan was recalculated using our in-house MC to obtain its LET$_d$ and BD distributions. The percentage gains in BD and LET$_d$ in plan 3 relative to plan 2 were quantified by:
\begin{equation} G_{BD} = 100\cdot \left(\frac{\overline{BD}_3}{\overline{BD}_2} - 1 \right) \quad , \quad G_{LET_{d}} = 100\cdot \left(\frac{\overline{LET}_{d3}}{\overline{LET}_{d2}}-1\right) \end{equation}
where $\overline{BD}_2$, $\overline{BD}_3$, $\overline{LET}_{d2}$, $\overline{LET}_{d3}$ are the mean target BD and LET$_d$ in plans 2 and 3. Similarly, the increase in mean PD in plan 3 relative to plan 2 is:
\begin{equation} G_{PD} = 100\cdot \left(\frac{\overline{PD}_3}{\overline{PD}_2} - 1 \right) \end{equation}
Ideally, it is desirable to have large $G_{BD}$ and $G_{LET_{d}}$ for a small $G_{PD}$, i.e. keep the target PD as close to the prescription dose as possible. The quantities
\begin{equation} R_D = \frac{G_{BD}}{G_{PD}}\quad , \quad R_{LET}=\frac{G_{LET_{d}}}{G_{PD}} \end{equation}
are used to assess the gain in BD and LET$_d$ relative to the increase in PD. Although the magnitude of $G_{BD}$ depends on the values of $a$ and $b$ in Equation 1, the optimization process does not. In other words, the choice of $a$ and $b$ does not affect the ability of the in-house TPS to converge to a biologically optimal plan.

For each patient, after optimization the BD distributions were generated using the RBE--LET$_d$ relations from references \cite{Carabe, Wedenberg, McNamara} and Equation 1. For each RBE model, the PD and BD Dose-Volume Histograms (DVHs) from plan 3 were compared with plan 2 to assess $G_{BD}$, $G_{PD}$, $R_D$ and OAR doses. To remove dependence on the coefficients $a$ and $b$, $R_D$ was calculated using the average $G_{BD}$ obtained from the Carabe, Wedenberg and McNamara models. Before calculating these metrics, all plans were normalized so that 98\% of the target gets the prescription dose. Comparisons were also made with plan 1 to gauge OAR PD sparing relative to standard IMRT treatment. Values of $G_{LET_{d}}$ and $R_{LET}$ were also calculated for each patient. In contrast to $G_{BD}$ and $R_D$, these are completely model-independent.
 
\subsection{Description of test cases}

This method cannot be successfully applied to all patients. Intuitively, the magnitude of $G_{BD}$ depends on the number of beams, their directions and target volume. Beam directions should be preferentially spread out isotropically around the target to facilitate LET$_d$ elevation. However, this somewhat defeats the purpose of proton therapy, and normal tissue irradiation should be avoided where possible. Furthermore, the contribution of any beam can be substantially reduced if an OAR is in its path. BD escalation in large-volume targets is expected to be more difficult to achieve, since low LET$_d$ track sections inevitably contribute to the dose.

In this study, four cases are discussed to illustrate the above points:
\begin{enumerate}[A.]
\item	Small thyroid tumor, OTV size: 94 cm$^3$
\item	Small thyroid tumor with OAR in the path of posterior beams, OTV size: 108 cm$^3$
\item	Large thyroid tumor, OTV size: 654 cm$^3$
\item	Small brain stem tumor, OTV size: 9 cm$^3$
\end{enumerate}
The field directions and target geometries in each case are shown in Figure 2 below. The prescription doses for cases A-D were 66, 60, 66 and 54 Gy, in 33, 50, 30 and 30 fractions, respectively. Cases B and C were included to demonstrate situations where our technique would not be applicable. In case B the left brachial plexus (shown as the yellow contour) is in the path of the posterior beams. Case D demonstrates the applicability of the technique to a site outside the neck region.

\subsection{Multi-GPU computing}
The initial PD and LET$_d$ map calculations, as well as the BD optimization, were carried out on a remote cluster \cite{bw} using 10 to 50 NVIDIA K20X cards, depending on the target volume. A multi-GPU system is mandatory due to the sheer scale of the initial map computations. In the optimization stage, the workload was also divided among a large number of GPUs. Fast inter-node communication was therefore essential for efficient data transfer. Further details of the multi-GPU implementation are given elsewhere \cite{Ma}.

\section{Results}

\begin{figure}
\includegraphics[width=\linewidth]{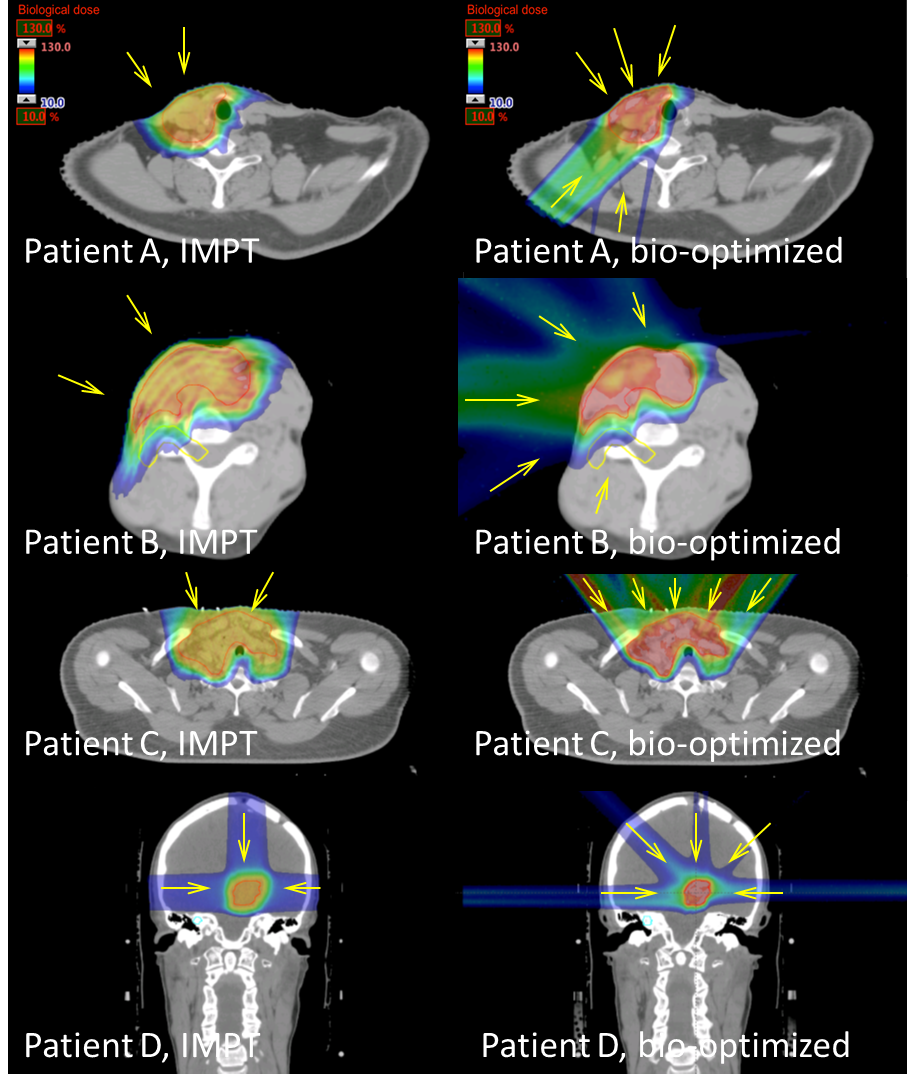}
\caption{The left column shows the BD distributions (relative to the prescription dose) from the MC recalculation of the 2-field Eclipse IMPT plans for patients A--D. The corresponding BD distributions from plans created with the in-house optimizer are shown on the right. Arrows indicate the beam directions. The targets are delineated in red. For patient B, the brachial plexus is delineated in yellow.}
\end{figure}

Figure 2 shows the BD distributions from the MC re-calculations of the conventional Eclipse IMPT plans (left) and our biologically optimized plans (right), for patients A to D.  The BD distributions in Figure 2 were generated using the McNamara model with $\alpha/\beta=3$ Gy (please see appendix for the LET$_d$ colorwash distribution). The values of $G_{PD}$, $G_{BD}$ from each publised RBE model \cite{Carabe, Wedenberg, McNamara} and their averages at $\alpha/\beta=3$ and 6 Gy, $R_D$, $G_{LET_d}$ and $R_{LET}$ for all cases are given in Table 1. 

Patients A and D had nearly 60\% increase in the mean target LET$_d$ and demonstrate the largest values of $R_{LET}$. The values of $G_{BD}$ and $R_{D}$ are model-dependent. For patients A and D with $\alpha/\beta=3$ Gy, raising the target PD by 5.1 and 8.6\% resulted in 11.3 and 17.3\% increase in target BD (i.e. relative gains of 2.21 and 2.02, respectively). Patients B and C had less favorable results, the gain in BD being marginal relative to the increase in PD. The gain in mean LET$_d$ relative to the PD increase was also much lower than for A and D. OAR DVHs for patients A (larynx and esophagus) and D (optic chiasm and pituitary gland) from the biologically optimized, standard IMPT and IMRT plans are shown in Figure 3. The target PD DVHs are shown in the appendix.

\begin{table}
\centering
\begin{tabular}{|c|c|c|c|c|}
\hline
Patient & A & B & C & D\\ \hline\hline
$G_{PD}$ & 5.1 & 8.2 & 13.6 & 8.6 \\\hline
$G_{BD}$ (Equation 1) & 12.3 & 10.9& 17.7& 18.2 \\\hline
$G_{BD}$ (Carabe et al, $\alpha/\beta=3$) & 12.2 & 9.1  & 17.9 & 18.6 \\\hline
$G_{BD}$ (Wedenberg et al, $\alpha/\beta=3$) &11.6& 8.4 & 17.1 & 17.7 \\\hline
$G_{BD}$ (McNamara et al, $\alpha/\beta=3$) & 10.0 & 8.6 & 16.4 & 15.7 \\\hline
$G_{BD}$ (average, $\alpha/\beta=3$) & 11.3 & 8.7 & 17.1 & 17.3 \\\hline
$G_{BD}$ (average, $\alpha/\beta=6$) & 10.4 & 8.9 & 16.8 & 15.7 \\\hline
$R_D$ ($\alpha/\beta=3$) & 2.209 & 1.061 & 1.260 & 2.016 \\\hline
$R_D$ ($\alpha/\beta=6$) & 2.029 & 1.082 & 1.236 & 1.823 \\\hline
$G_{LET_{d}}$ & 58.3 & 14.3 & 36.3 & 59.4 \\\hline
$R_{LET}$ & 11.438 & 1.742 & 2.674 & 6.904 \\
\hline
\end{tabular}
\caption[] {Post bio-optimization metrics (defined in Equations 2--4) for cases A--D. }
\end{table}

\begin{figure}
\includegraphics[width=\linewidth]{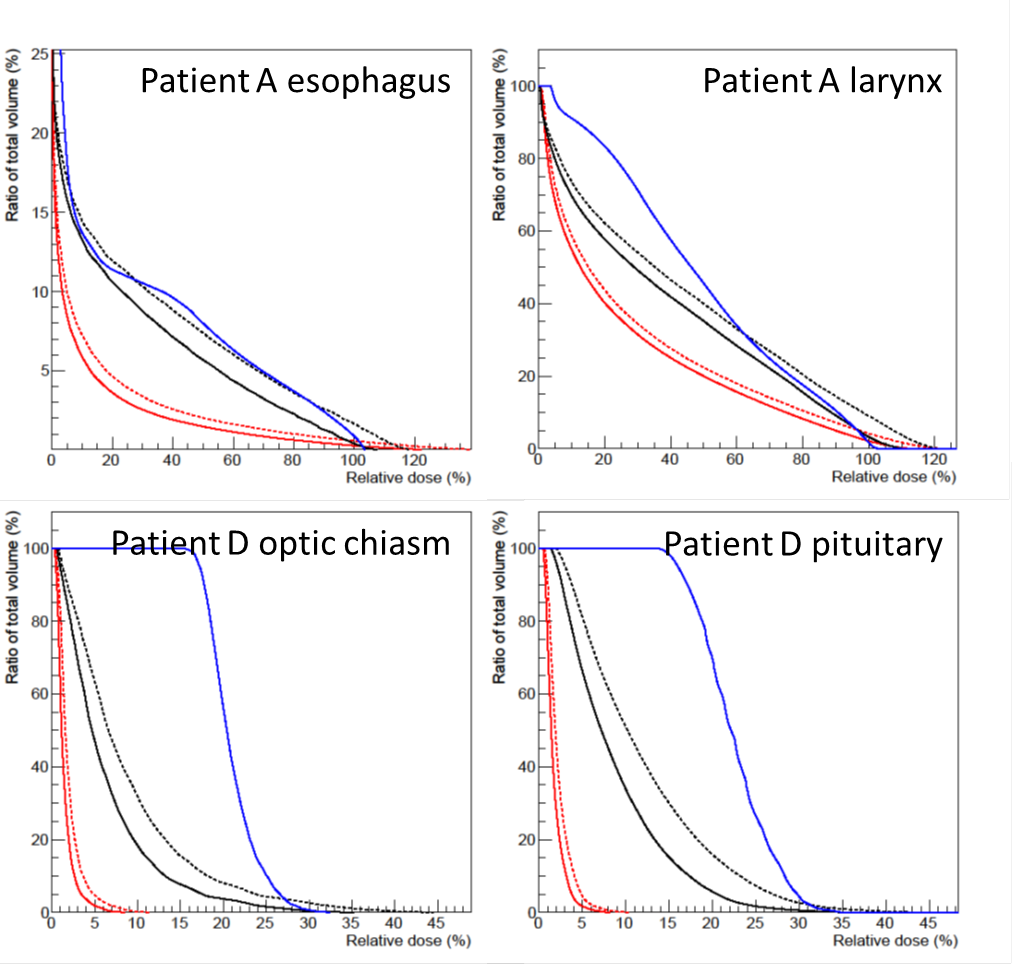}
\caption{Top row: larynx and esophagus DVHs for patient A. Bottom row: optic chiasm and pituitary gland DVHs for patient B. Solid red, black and blue lines are the PD DVHs from the biologically optimized, IMPT and IMRT plans, respectively. The proton PD has been scaled by an average RBE value of 1.1 for comparison with IMRT. The dashed lines are the BD DVHs for the proton plans (calculated using the McNamara model with $\alpha/\beta=3$ Gy).}
\end{figure}

\subsection{Calculation time}
The plan calculation time was dominated by the initial PD and LET$_d$ map calculations. The optimization process itself typically took less than 17 minutes on the cluster. Table 2 gives the number of beam spots covering the STV for each patient, as well as the total number of protons in the spot maps. The total dose map processing (including the actual map calculation on the GPU, GPU-CPU transfers, initialization and file output) and optimization times in seconds are also given. The number of GPUs used during the optimization stage is shown in the last row.

\begin{table}
\centering
\begin{tabular}{|c|c|c|c|c|}
\hline
Patient & A & B & C & D\\ \hline\hline
Target volume (cc) & 94 & 108 &654 & 9 \\\hline
Number of spots & 26184 & 55531& 121582&4255 \\\hline
Number of protons in dose map &$2.62\times10^9$&$5.56\times10^9$&$1.22\times10^{10}$&$4.25\times10^8$\\\hline
Total GPU time to create dose map (s) &18086.4 & 33041.4 & 81296.7 & 3497.4 \\\hline
Average time per proton history ($\mu$s) & 6.9 & 5.9 & 6.7 & 8.2 \\\hline
Optimization time (minutes) & 10.6 & 4.7 & 16.5 & 1.1 \\\hline
No. of GPUs used in optimization step & 25 & 10 & 50 & 10 \\
\hline
\end{tabular}
\caption[] {The total number of spots and proton histories in the initial dose and LET$_d$ map calculations for each patient are shown in the second and third rows, respectively. The net GPU processing times, average time per proton history, optimization time and number of GPUs used during optimization are shown in rows 4, 5, 6 and 7, respectively. All calculations were performed on NVIDIA K20X cards. }
\end{table}

\section{Discussion}
BD optimization by MC-based inverse planning has been demonstrated for two patients with small tumor targets (A and D). For a modest rise in mean target dose, roughly twice as much increase in BD was obtained. $R_D$ and $R_{LET}$ are a rough guide to the extent of the BD escalation `quality'. A high value of $R_D$ means that not much additional PD is required to significantly boost the target BD, which implies that the target dose can stay relatively close to prescription. For plans with high $R_D$ and $R_{LET}$, the BD gain is a consequence of the successful and deliberate placement of high-LET track ends inside the targets. 

Compared to conventional IMPT planning, more beams were required to elevate LET$_d$ in the target. As observed in Figure 2, the beams were not spread uniformly around the target in order to spare normal tissue. However, a higher degree of BD escalation could have been achieved by irradiating the opposite areas. A five-beam arrangement was used in this work because splitting the beams (but keeping the same anatomical avoidance regions) does not result in significantly larger $R_D$ and $R_{LET}$ values. For a small-target patient, the BD gain of a 10-field plan was found to be comparable to that of a 5-field plan.

The limitations of the methods are illustrated by cases B and C. As shown in case C, achieving BD escalation in large-volume targets is far more challenging because of the unavoidable contribution of low-LET track sections to PD. Case B has comparable volume to A, but the presence of an OAR (the left brachial plexus) in the path of the posterior beams considerably diminished their contribution. As a result, the remaining field directions ended up being rather similar to those in the IMPT plan, and no significant BD boost was achieved.

Figure 3 shows that dose sparing to critical structures in the biologically optimized plans is improved compared to both Eclipse IMRT and IMPT plans. In contrast to Eclipse, the in-house optimization algorithm continuously drove down the dose to a given OAR even when user-defined constraints on it were met. To achieve this, the user-defined upper constraints and their weightings were automatically adjusted during each iteration step \cite{Ma}.

It might be thought that the BD gain relative to standard IMPT planning was mostly due to the additional fields, and that a 5-field IMPT plan would demonstrate a similar level of BD gain. In fact, BD escalation cannot be deliberately achieved if the optimizer is not explicitly instructed to do so. To illustrate this, a 5-field standard IMPT plan was created in Eclipse for patient D, using the same beam angles, constraints and objectives (except the BD objective, which cannot be applied) as the biologically optimized plan. Figure 4 shows the LET$_d$ colorwash distributions for the 5-field IMPT and biologically optimized plans in an axial cut. The central plot shows the LET$_d$ profile across the target. Although the target PD objectives and OAR constraints were met by both plans, the LET$_d$ distributions are very different from each other. The two peaks in the red curve indicate that Eclipse placed high-LET proton track ends near the target edges. Conversely, the biologically optimized plan was constructed such that protons stopped preferentially within the target, resulting in BD escalation. Relative to the 3 field conventional IMPT plan, the values of $G_{PD}$, $G_{BD}$ and $R_D$ ($\alpha/\beta=3$ Gy) obtained for the 5-field Eclipse IMPT plan were 1.8\%, 2.1\% and 1.17, showing no evidence for BD escalation.

\begin{figure}
\includegraphics[width=\linewidth]{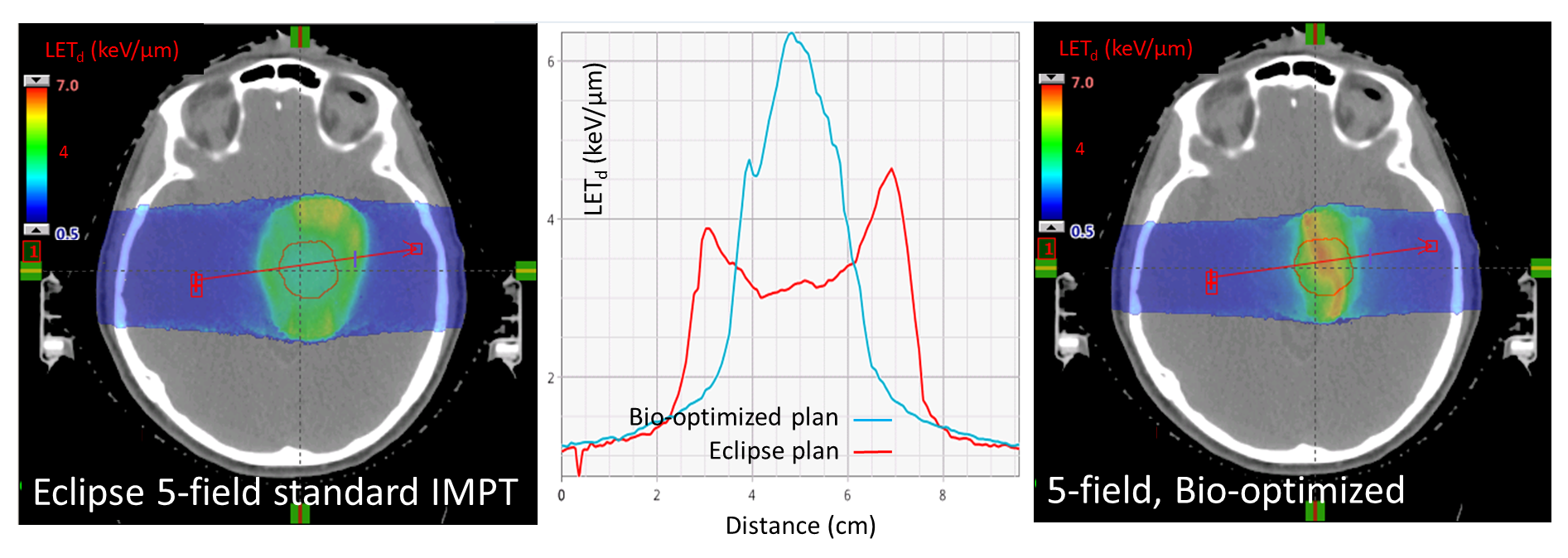}
\caption{LET$_d$ colorwash distributions for a standard 5-field IMPT plan (left) and our biologically-optimized plan (right) for patient D. The LET$_d$ for voxels receiving less than 1\% of the maximum dose is not shown. The central plot shows the LET$_d$ profile across the target. The same beam angles, objectives and constraints were used in both plans (see text).}
\end{figure}

It should be pointed out that although our LET$_d$ predictions are in very good agreement in TOPAS, recent work \cite{Carabelet, Granville, Guan} has shown that the default LET$_{d}$ scoring method in early TOPAS versions (including beta-6, which we used in this work) is biased. To resolve this, the LET$_d$ scoring method adopted in our MC will be modified to use tabulated unrestricted stopping power tables, based on recommendations from references \cite{Carabelet, Granville, Guan}. In the near future, we also plan to update our optimizer to use a linear-quadratic RBE model \cite{McNamara} instead of Equation 1. The use of a linear RBE model in this feasibility study was purely out of simplicity, and this update is not expected to impact our ability to achieve BD escalation.

As mentioned previously, a multi-GPU system is necessary to perform the calculations within a clinically acceptable timeframe. Compared to a K20X, a Titan X card can process a given dose calculation in around half the time. From Table 2, assuming the processing time to scale with the number of GPUs, a system consisting of 24 NVIDIA Titan X cards can complete the calculations for case A in under 30 minutes.

\subsection{Feasibility trial on LET$_d$ painting}
A feasibility trial on proton BD optimization is planned to start at our institution in 2016.  This trial is geared towards patients with: (a) anaplastic thyroid cancers, (b) salivary gland tumors, (c) recurrent squamous cell carcinomas of the head and neck, and (d) recurrent glioblastomas. A total of 24 patients will be enrolled, with 6 from each tumor category type. The tumor target size will be critical in selecting patients. The treatment planning calculations will be carried out on a small local cluster of 24 GPU cards, using the method described above. Candidates will be planned using both standard IMPT and biological planning, and the treatment modality will be chosen at the physician’s discretion. The requirements for treatment using the biologically optimal plan include large $G_{BD}$, $G_{LET_{d}}$, $R_D$ and $R_{LET}$ values. 

\section{Conclusions}
By taking advantage of GPU acceleration, the BD in small tumors was successfully escalated via MC-based inverse planning optimization. In the examples given, compared to standard IMPT planning, a modest increase in the target PD significantly raised the mean LET$_d$ and can result in twice as much increase in the BD. The estimated BD increase depends on the LET$_d$ to BD conversion model. The applicability of the technique was discussed. To summarize, its main advantages are:
\begin{itemize}
\item Accurate MC-based dose and LET$_d$ calculation
\item Inverse planning optimization
\item Clinically applicable
\end{itemize}
The main limitations are:
\begin{itemize}
\item Irradiation to normal tissue due to additional fields
\item Not applicable to all cases, especially large target volumes
\item Expected to be less robust than standard IMPT
\end{itemize}
The ability to produce such plans within a clinically acceptable timeframe opens exciting new prospects for proton therapy.

\section{Acknowledgements}
This work was partially funded by Varian Medical Systems, Inc.

\appendix
\section{Supplemental Materials}
\begin{figure}
\includegraphics[width=13cm]{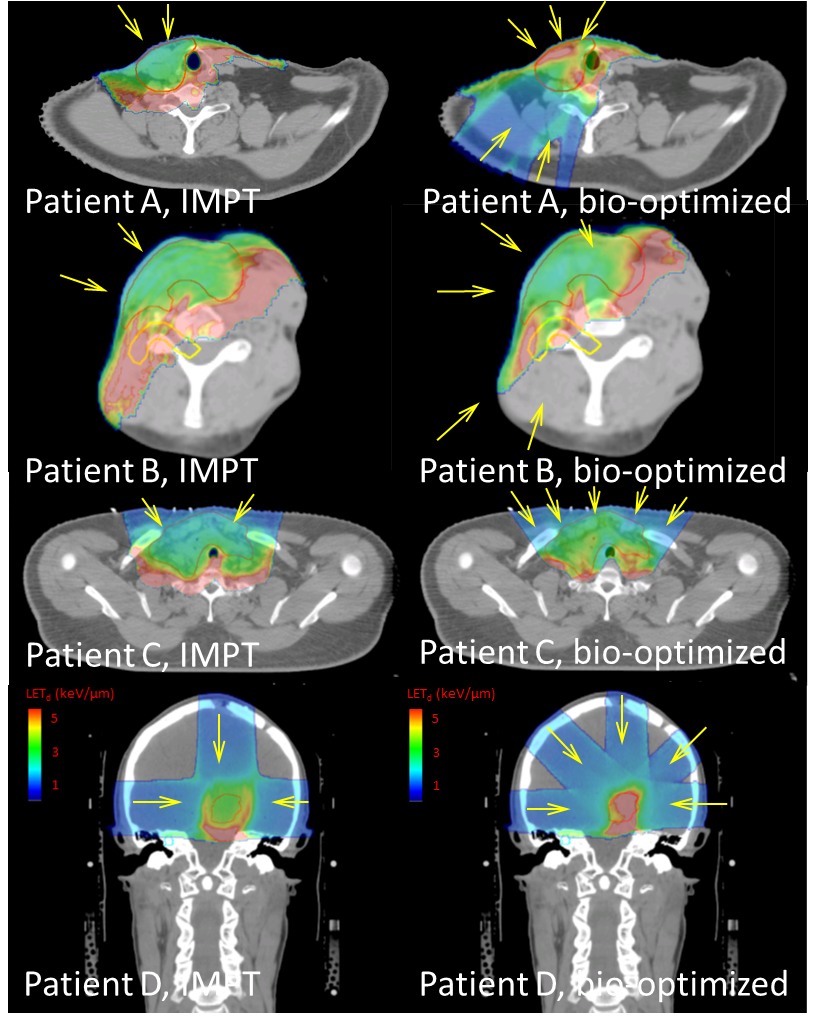}
\caption{The left column shows the LET$_d$ distributions from the MC recalculations of the Eclipse IMPT plans for aptients A--D. The LET$_d$ distributions for the biologically optimized plans are shown on the right, in the same slices as Figure 2 of the paper. Targets are delineated in red. For patients A and D, the mean target LET$_d$ is increased by nearly 60\%.}
\end{figure}

\begin{figure}
\includegraphics[width=\linewidth]{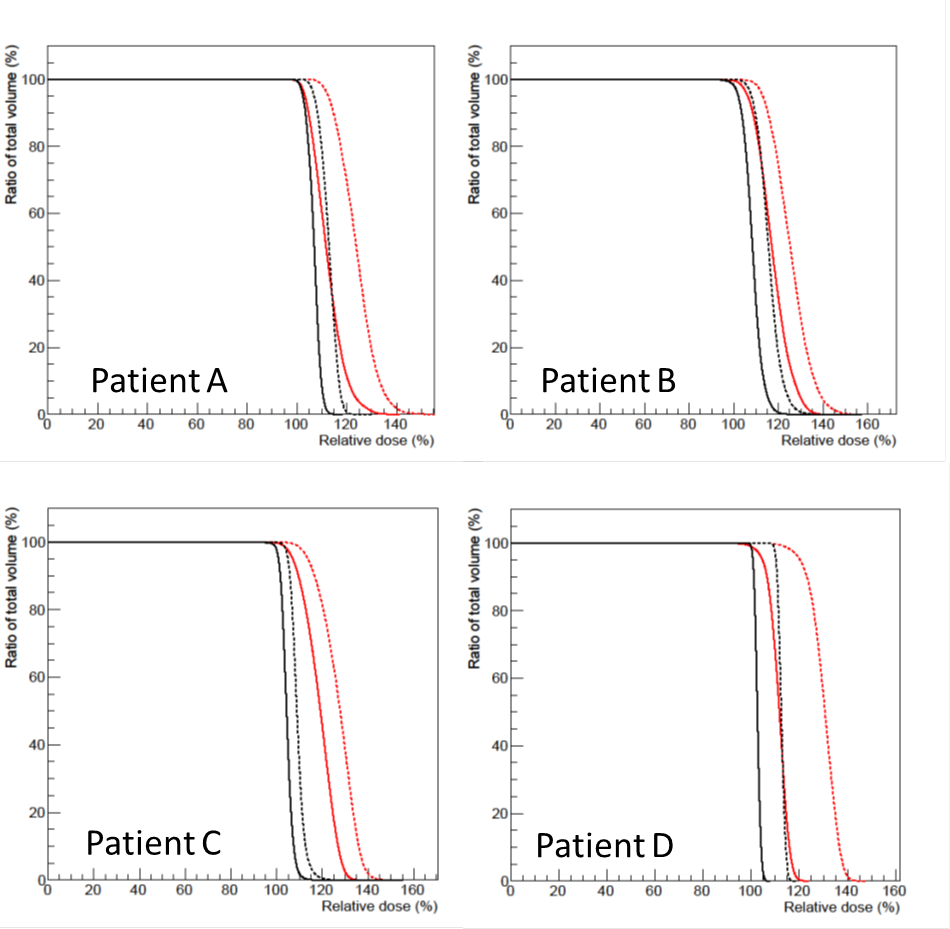}
\caption{Target DVHs for patients A--D. Solid and dashed curves are the PD and BD DVHs, respectively. Black curves are from the MC re-calculations of the Eclipse IMPT plans, while red curves are from the biologically optimized plans.}
\end{figure}

\end{document}